\documentclass[a4paper,10pt]{article}
\usepackage[utf8]{inputenc}
\usepackage[english]{babel}
\usepackage{graphicx}
\usepackage[table]{xcolor}
\usepackage{amsfonts}
\usepackage{units}
\usepackage{amsmath,amssymb,verbatim,float,tabularx,epic}
\usepackage{epsfig}
\usepackage{subfigure}
\usepackage{listings}
\usepackage{rotating}
\usepackage{numprint}
\usepackage{url}

\title{Solving a real-life large-scale\\energy management problem}
\author{Steffen Godskesen \and Thomas Sejr Jensen \and Niels Kjeldsen \and Rune Larsen}

\begin{document}

\maketitle

\begin{abstract}
This paper introduces a three-phase heuristic approach for a large-scale energy management and maintenance scheduling problem. The problem is concerned with scheduling maintenance and refueling for nuclear power plants up to five years into the future, while handling a number of scenarios for future demand and prices. The goal is to minimize the expected total production costs. The first phase of the heuristic solves a simplified constraint programming model of the problem, the second performs a local search, and the third handles overproduction in a greedy fashion.

This work was initiated in the context of the ROADEF/EURO Challenge 2010, a competition organized jointly by the French Operational Research and Decision Support Society, the European Operational Research Society, and the European utility company \'{E}lectricit\'{e} de France. In the concluding phase of the competition our team ranked second in the junior category and sixth overall.

After correcting an implementation bug in the program that was submitted for evaluation, our heuristic solves all ten real-life instances, and the solutions obtained are all within 2.45\% of the currently best known solutions. The results given here would have ranked first in the original competition.
\end{abstract}

\section{Introduction}
\'{E}lectricit\'{e} de France is the main supplier of electricity in France. The majority of the electricity is produced by thermal --- and in particular nuclear --- power plants. There are two types of thermal power plants in their portfolio: \emph{type 1 plants} which can be supplied with fuel continuously and without interrupting the production and \emph{type 2 plants} (nuclear power plants) which must be taken offline for refueling at regular intervals. While type 1 plants are more flexible than type 2 plants, the production cost incurred per unit of electricity is larger than for type 2 plants. The process of taking a power plant offline for maintenance is called an \emph{outage}. Scheduling outages for type 2 plants should be done such that the estimated demand for electricity is satisfied at the lowest possible cost. A schedule for outages must satisfy a large number of constraints due to e.g. limited resources and safety considerations.

Some of the constraints are due to limits on fuel levels in connection with each outage: There is a maximum feasible fuel level for the power plant, a maximal allowed fuel level when a plant is taken offline, and a minimum refueling amount for each outage. When planning future production, we have to decide the timing of each outage, refuel amounts, and production levels.

The total production of electricity is not allowed to exceed the demand, and therefore type 2 plants sometimes produce at less than their maximum production level. This is called \emph{modulation}, and due to technical reasons there is a limit on the allowed modulation for each plant for each \emph{production campaign} --- the period between two outages.

When a type 2 plant is taken offline for refueling and maintenance, the electricity must instead be produced by alternative sources. This can either be done by one of the other type 2 plants or by the more expensive type 1 power plants. The future electricity demand and the price of fuel for production on type 1 plants is uncertain. This uncertainty should be taken into account in the planning of outages for type 2 plants. This is handled here by minimizing the expected future production cost.

\subsection{The ROADEF/EURO Challenge 2010}
The described problem was the subject of the ROADEF/EURO Challenge 2010, which ran from July 2009 through June 2010. In total 44 teams from 25 countries signed up for the challenge, of these 21 qualified for the final round, and 16 submitted a program for the final evaluation. The submitted programs were evaluated on ten problem instances, five known and five unknown. For each instance the time limit imposed to the program was one hour. In the concluding phase of the competition our team ranked second in the junior category and sixth overall. A complete description of the competition and evaluation rules can be found in \cite{roadefWeb}.

The problem instances from the challenge have up to 75 type 2 plants, up to 120 scenarios for future prices and demand, and up to 5817 discrete time steps. Outages always start at the beginning of a week and since the time horizon is about five years there can be up to about 260 possible outage start dates. All in all this leads to a large-scale energy management problem.

\begin{figure} [tb]
	\centering
		\includegraphics[width=\textwidth, trim=0 30 0 60]{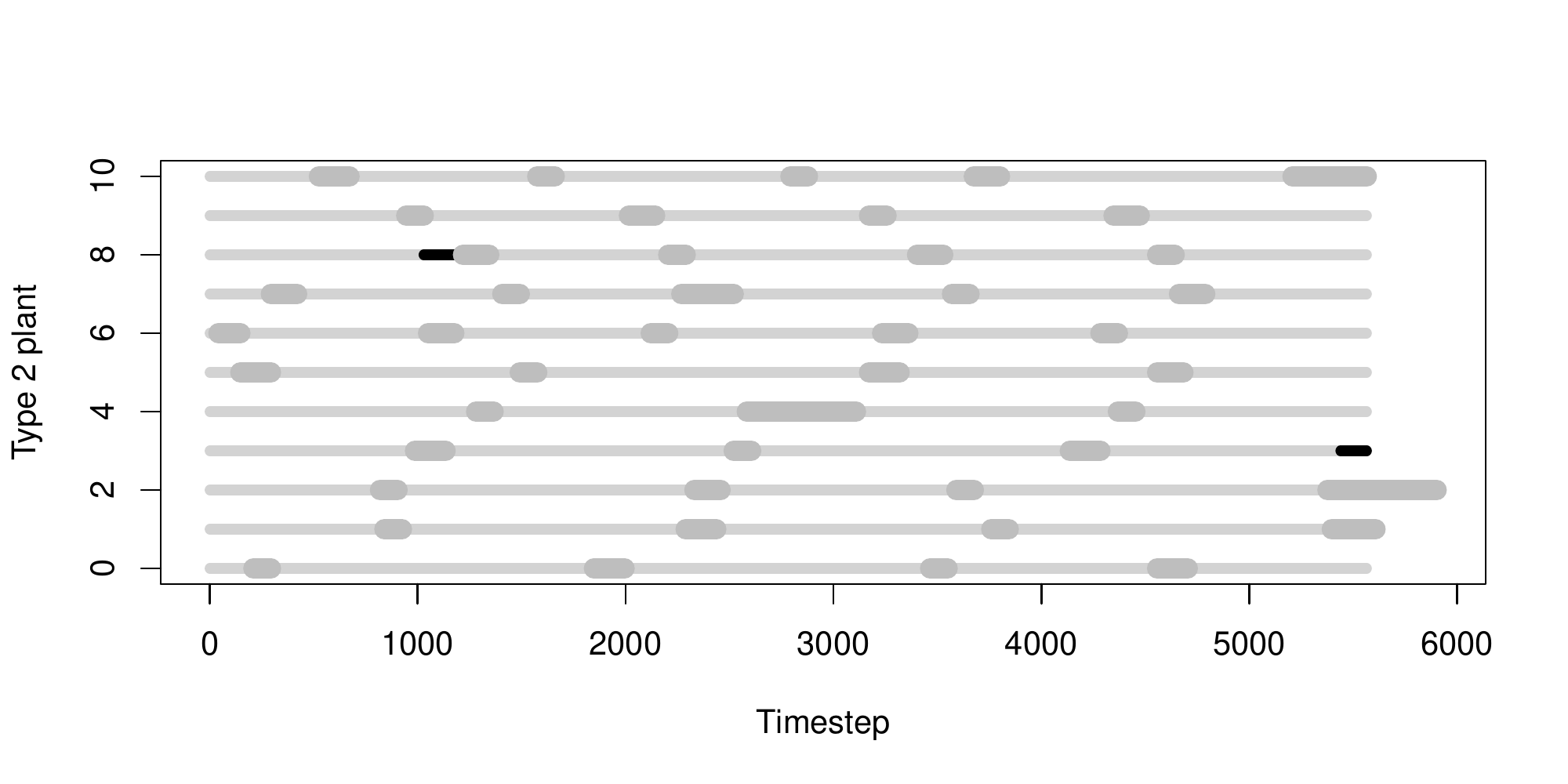}
		\caption{A sample outages schedule for 11 plants. For all time steps the state of each plant is given. A light grey line is used for time steps with production, a wide dark grey for outages, and a black is when the plant is out of fuel.}
	\label{fig:outageschedule}
\end{figure}

A solution to the problem can be divided in two parts: a scheduling part and a production planning part. To give an indication of what a schedule for outages looks like see Figure \ref{fig:outageschedule} which shows a schedule for 11 type 2 plants.

\begin{figure} [tb]
	\centering
		\includegraphics[width=\textwidth, trim=0 30 0 60]{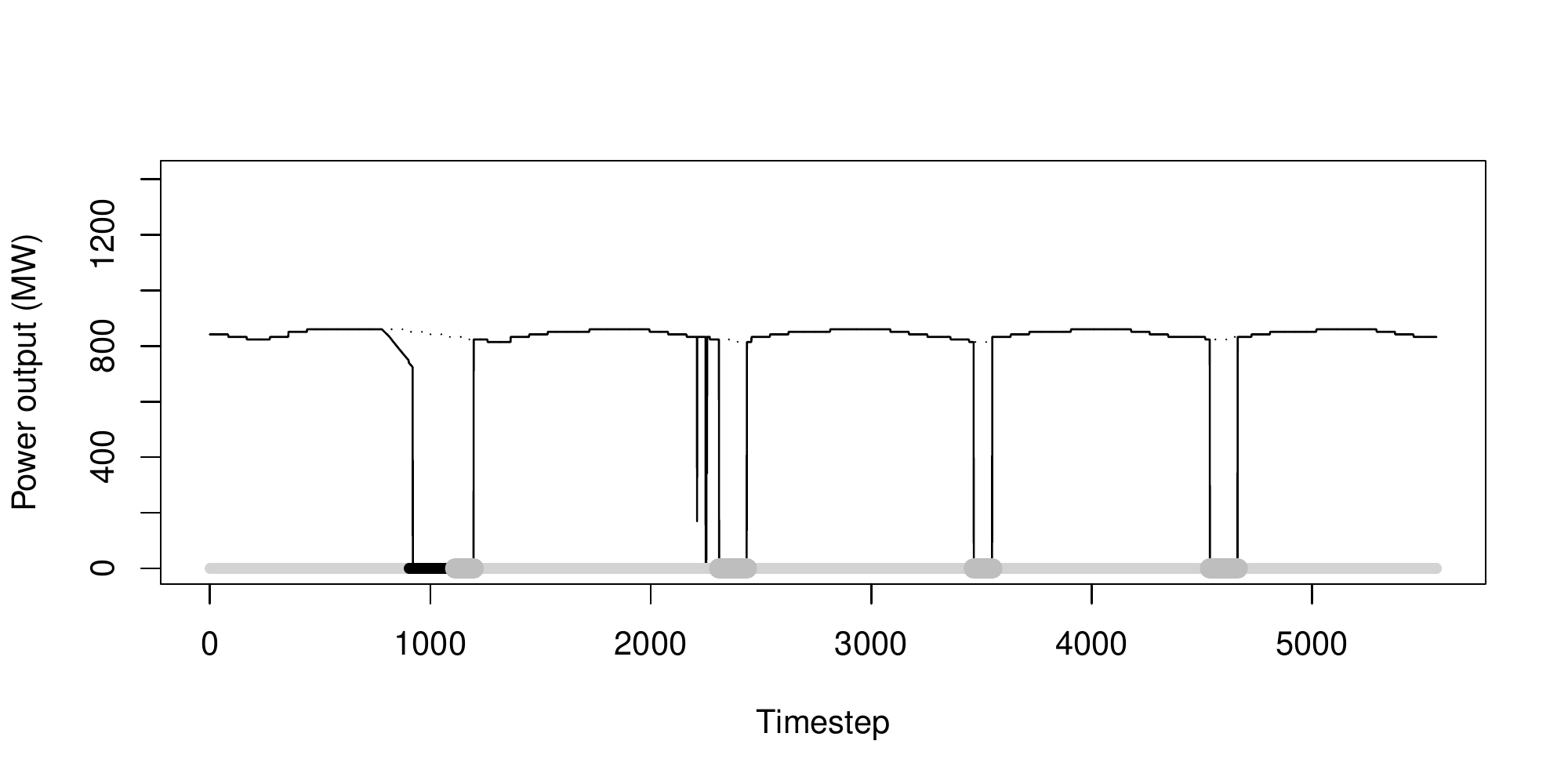}
		\caption{The production level over time for plant 8 from Figure \ref{fig:outageschedule}. The full line is the current production level and the dashed line the maximum allowed production level.}
	\label{fig:production}
\end{figure}

\begin{figure} [tb]
	\centering
		\includegraphics[width=\textwidth, trim=0 30 0 60]{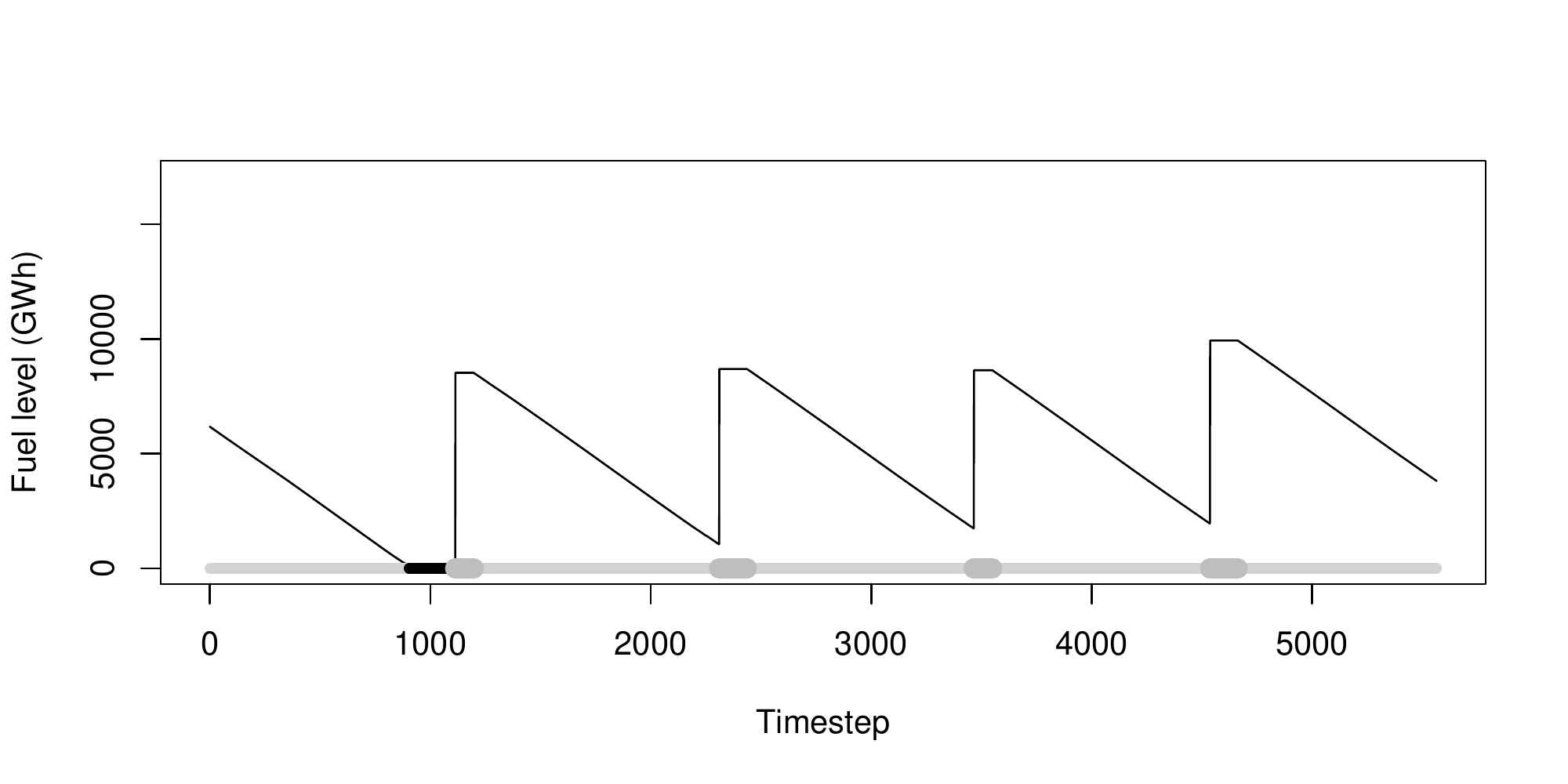}
		\caption{The fuel level over time for plant 8 from Figure \ref{fig:outageschedule}.}
	\label{fig:fuel}
\end{figure}

Figures \ref{fig:production} and \ref{fig:fuel} show production level and fuel level, respectively, for type 2 plant number 8 over time. Note the sudden decrease in production around time step \numprint{2200}; this is modulation to ensure that there is no overproduction in the specific timestep.

The following factors indicate that the problem is hard. First of all, it is NP-hard in the strong sense, as we prove in Section \ref{NP-hardness}. Furthermore, the problem instances are large, as a single problem instance takes up to 262 megabytes of harddisk space and contains more than $50 \cdot 10^6$ continuous decision variables just to represent production levels for every plant, time step, and scenario. Finally, there is a large number of constraints on production levels, fuel levels, refueling levels, and on the scheduling of outages.

The program we submitted contained an implementation bug which resulted in infeasible solutions for two of the five unknown instances (feasible solutions were found for all other instances). 
After correcting program we are able to solve all instances, and the solutions are actually better than those of the winning team (note that most teams are probably able to improve on their program after all instances are made publicly available). 

\subsection{Related work}
A problem similar to the one studied here has been considered in \cite{fourcade1997optimizing} by Fourcade et al.\ in 1997. They consider roughly the same scheduling problem as is handled here and formulate a mixed integer programming model. In their model there is no decision variable concerning refueling amounts; this decision is instead handled as a predefined fixed amount. There is also no uncertainty of future demand or prices, and the demand is given per week, in contrast to the competition where the electricity demand is given per time step, i.e., their discretization is more coarse-grained.

They are able to solve small problems of up to 20 nuclear power plants with a MIP solver. The authors also attempted to tackle a model with 40 power plants, which yields a feasible solution after more than eight hours of computational time, but with a significant gap to the LP lower bound. They report that this model is almost the size of the complete model of France which has 54 nuclear plants. The performance of MIP solvers and hardware have increased significantly since 1997, so modelling that particular problem as a mixed integer programming problem might be feasible with state of the art solvers.

Besides the work by Fourcade et al.\ very little is published on the topic. Nuclear maintenance and refueling is mentioned by Dunning et al.\ in \cite{dunning2001new}, but the problem considered is to minimize the environmental impact.


The scheduling part of this problem is quite similar to the Resource Constrained Project Scheduling Problem (RCPSP), see e.g.\ \cite{scheduling_with_time_windows}. In both problems, activities (in this case outages) have to be be scheduled subject to temporal constraints and limited resources. However, the problem at hand includes several constraints not found in common variants of RCPSP, such as temporal constraints that are disjunctive and only apply if a pair of activities is scheduled in a specified interval (as described in Section \ref{constraints}).

Setting production levels for power plants is treated in the literature under the term `economic dispatch' --- i.e., the problem of dispatching units to producing power in an economic way to minimize production costs. While many different settings are considered, see for example \cite{chowdhury2002review} by Chowdhury and Rahman, there are new features in the production planning for nuclear power plants. The new features concern special bounds on production levels when the fuel level is low, which leads to nonlinear constraints. When a type 2 plant's fuel level drops below a given treshold, a decreasing power production level is imposed. Without this constraint the production planning could be solved with linear programming.

\subsection{Our contribution}
The purpose of this paper is to introduce a three-phase heuristic approach for the problem sketched above. A constraint programming (CP) model of the scheduling problem with approximated constraints for production levels and fuel consumption gives us a feasible maintenance schedule. From this first feasible solution we apply a local search algorithm based on a simple neighborhood structure. Two essential components in the local search algorithm are: a fast feasibility check and a fast, but approximative, evaluation of the change in solution cost. To guide the local search we embed it in a simulated annealing metaheuristic. In the third and final phase we use a greedy algorithm to remove overproduction, if any. The primary task of the algorithm is to make the given solution feasible, and consequently it may not be optimal.

\subsection{Overview}
The paper is organized as follows. Section \ref{problem_description} gives a formal decription of the optimization problem as well as a complexity proof. In Section \ref{initial_solution_construction} we describe how to obtain a feasible solution using a CP model and a heuristic for production planning. Our simulated annealing algorithm is described in Section \ref{improvement_by_local_search}. Section \ref{modulation} describes how to handle overproduction. Computational analysis and results are the topic of Section \ref{computational_analysis}. Finally, we conclude and give directions for further research in Section \ref{conclusion}.



\section{Problem description} \label{problem_description}
Each type 2 plant goes through a number of \emph{cycles}. A cycle is composed of an \emph{outage} followed by a \emph{production campaign}. During an outage the plant cannot produce electricity because of maintenance and reloading of fuel. During a production campaign the plant is able to produce electricity. Having type 2 plants that produce at less than maximum capacity leads to wear of the equipment involved and should thus be avoided if possible. The difference between maximum capacity and actual production is called \emph{modulation}.

The demand for electricity is not known with certainty at the time of planning. This stochasticity is dealt with by introducing a number of \emph{scenarios}, each of which represents a realistic future demand profile for the planning horizon discretisized into a number of time steps. Optimizing for several realistic scenarios instead of just one generally leads to more robust plans.

Decisions concerning scheduling of outages and refueling amounts are shared by all scenarios, but production levels are determined for each individual scenario. This creates a dependency between scenarios, since the outage schedule and refueling amounts must be feasible with respect to every scenario's production plan.

A \emph{production plan} specifies the production level of each plant for every time step and every scenario. Furthermore, a \emph{maintenance plan} specifies when outages of type 2 plants are scheduled and the amount of fuel to reload at each outage. The objective is to satisfy the demand for electricity at the lowest average cost over all scenarios.

The cost must be minimized while satisfying a number of constraints, which can be divided into four categories: i) bounds on production levels, ii) bounds on refueling amounts, iii) different kinds of temporal constraints on the scheduling of outages, including constraints involving outages for different type 2 plants, and iv) bounds on the outages' simultaneous use of limited resources.

\subsection{Decision variables and bounds}
We use $s=0, \mathellipsis , S-1$ to index scenarios, $t=0, \mathellipsis , T-1$ to index time steps, $h=0, \mathellipsis , H-1$ to index weeks, $j=0, \mathellipsis , J-1$ to index type 1 plants, $i=0, \mathellipsis , I-1$ to index type 2 plants, and $k=0, \mathellipsis , K-1$ to index cycles. A week consists of a number of time steps, i.e., two different discretizations of the planning horizon are used. This is because outages are scheduled on a weekly basis, whereas a higher resolution is required when determining productions levels. The length of a time step in hours is denoted by $D$ (all time steps have the same length).

As in the original problem formulation \cite{roadefWeb}, we index decision variables using parentheses in order to distinguish them from parameters in the model. Let $p(\ell,t,s) \geq 0$ denote the production of plant $\ell$ (which may be of type 1 or 2) at time step $t$ in scenario $s$. 

The length of outage $k$ for type 2 plant $i$ is denoted by $DA_{i,k}$. Let $ha(i,k) \in \mathbb{Z}$ denote the week that the $k$'th outage for type 2 plant $i$ starts, and $TO_{i,k}$ and $TA_{i,k}$ denote the lower and upper bound, respectively, on $ha(i,k)$. Then we have
\begin{equation} \label{outage_bounds}
TO_{i,k} \leq ha(i,k) \leq TA_{i,k} . 
\end{equation}
The bounds $TO_{i,k}$ and $TA_{i,k}$ may be undefined, in which case the corresponding inequality is trivially satisfied. If the upper bound is undefined for some outage, the outage does not have to be scheduled. If outage $k$ for plant $i$ is not scheduled then $ha(i,k)=-1$ and constraint \eqref{outage_bounds} is not enforced. Outage $k+1$ for some plant cannot start before outage $k$ for the same plant is finished.

The amount of fuel reloaded at type 2 plant $i$ in outage $k$ is denoted by $r(i,k) \geq 0$ and must satisfy \eqref{reload_bounds} if $k$ is scheduled
\begin{equation} \label{reload_bounds}
RMIN_{i,k} \leq r(i,k) \leq RMAX_{i,k} , 
\end{equation}
where the bounds $RMIN_{i,k}$ and $RMAX_{i,k}$ are input data. If $k$ is not scheduled, then $r(i,k) = 0$.

\subsection{Auxiliary variables}
In addition to the decision variables there is a number of auxiliary variables which can be derived from the decision variables and thus do not increase the size of the solution space.

The set of time steps composing the $k$'th outage of type 2 plant $i$ is denoted by $ea(i,k)$, and the set of time steps composing the subsequent production campaign is denoted by $ec(i,k)$. For any $k$, the production $p(i,t,s)$ of plant $i$ must be zero for every $t \in ea(i,k)$ in every scenario $s$.

The fuel stock of type 2 plant $i$ at time step $t$ in scenario $s$ is denoted by $x(i,t,s) \geq 0$. The initial fuel level of $i$ (at time step 0) $XI_i$ is specified in the input data. During a production campaign for plant $i$ the decrease in fuel level from time step $t$ to $t+1$ in scenario $s$ equals the production multiplied by the length of a time step $D$
\begin{equation}
x(i,t+1, s) = x(i,t,s) - p(i,t,s) \times D.
\end{equation}
During an outage the fuel level at type 2 plant $i$ increases because of refueling. Due to technical reasons the new fuel level is not simply the sum of the old fuel level and the amount reloaded. Formally, if $t$ is the first time step in outage $k$ for plant $i$, then the new fuel level for $i$ in scenario $s$ is computed using
\begin{equation} \label{fuel_reload}
x(i, t+1, s) = Q_{i,k} \times x(i,t,s) + r(i,k) + Q'_{i,k} ,
\end{equation}
where $Q_{i,k} < 1$ and $Q'_{i,k}$ are input data\footnote{Equation \eqref{fuel_reload} is a simplification of Equation (CT10) in the original model defined by ROADEF, but the two formulas are equivalent when appropriate values for $Q_{i,k}$ and $Q'_{i,k}$ are used.}

\subsection{Constraints} \label{constraints}
The constraints can be divided into three groups, namely production level constraints, fuel level constraints, and scheduling constraints.

\paragraph{Production level constraints}
Let $DEM^{t,s}$ denote the \emph{demand} at time step $t$ in scenario $s$. The total production must equal the demand in every scenario and every time step
\begin{equation} \label{demand_constraint}
\forall s,t: \sum_{j=0}^{J-1} p(j,t,s) + \sum_{i=0}^{I-1} p(i,t,s) = DEM^{t,s}.
\end{equation}
Let $PMIN_j^{t,s}$ and $PMAX_j^{t,s}$ denote the minimum and maximum, respectively, \emph{allowed production} of type 1 plant $j$ at time step $t$ in scenario $s$, then
\begin{equation}
\forall s,t: PMIN_j^{t,s} \leq p(j,t,s) \leq PMAX_j^{t,s}.
\end{equation}
The bounds on production for a type 2 plant are more complex, since the depend on the current fuel stock of the plant. If the fuel level is above a threshold $BO_{i,k}$ that depends on the production campaign $k$, then the production is bounded from above by $PMAX_i^t$
\begin{equation}
\forall s,t,i,k: t \in ec(i,k) \wedge x(i,t,s) \geq BO_{i,k} \Rightarrow  0 \leq p(i,t,s) \leq PMAX_i^t .
\end{equation}
As long as the fuel level is above the threshold, there is no lower bound on the production of type 2 plants in each individual time step, but modulation is undesirable and therefore there is an upper bound $MMAX_{i,k}$ on the accumulated modulation of plant $i$ in each production campaign $k$
\begin{equation}\label{eq:maxmod}
\forall s,i,k: \sum_{\substack{t \in ec(i,k) \wedge \\ x(i,t,s) \geq BO_{i,k}}} (PMAX_i^t - p(i,t,s)) \times D \leq MMAX_{i,k} .
\end{equation}
If the fuel level is below the threshold, the upper bound decreases and a lower bound is also enforced. This is referred to as the \emph{declining power profile}. How much the upper bound decreases for type 2 plant $i$ in production campaign $k$ is specified by a function $PB_{i,k}$ which maps fuel level to a real number between zero and one. Formally, for all $s,t,i,k$, if $t \in ec(i,k)$ and $x(i,t,s) < BO_{i,k}$, then the production must lie in a small interval centered around $P_x$
\begin{equation} 
P_x = PB_{i,k}(x(i,t,s)) \times PMAX_i^t , 
\end{equation}
\begin{equation}
\label{power_profile_enough_fuel}
(1-\epsilon) \times P_x \leq p(i,t,s) \leq (1+\epsilon) \times P_x .
\end{equation}
However, if the plant will run out of fuel if it produces at $P_x$, it cannot produce at all. Thus, \eqref{power_profile_enough_fuel} applies only if inequality \eqref{enough_fuel} holds. If \eqref{enough_fuel} does not hold, $p(i,t,s)$ must be zero
\begin{equation} \label{enough_fuel}
x(i,t,s) \geq P_x \times D . 
\end{equation}

\paragraph{Fuel level constraints}
There are upper bounds on the \emph{fuel level} before and after a type 2 plant outage. Let $AMAX_{i,k}$ denote the upper bound on the fuel level at the time when outage $k$ for plant $i$ starts and $SMAX_{i,k}$ the upper bound on the fuel level after outage $k$ for plant $i$. If the $k$'th outage for plant $i$ starts at time step $t$ in some scenario $s$, inequality \eqref{amax} and \eqref{smax} must hold
\begin{equation} \label{amax}
x(i,t,s) \leq AMAX_{i,k} , 
\end{equation}
\begin{equation} \label{smax}
x(i,t+1,s) \leq SMAX_{i,k} .
\end{equation}

\paragraph{Scheduling constraints}
There are disjunctive \emph{temporal constraints} between outages. If a specified pair of outages $(i,k)$ and $(i', k')$ is scheduled such that they both intersect a specified interval (this interval may be the entire planning horizon), then constraint \eqref{min_separation} must be satisfied 
\begin{equation} \label{min_separation}
ha(i,k) - ha(i', k') \geq Se \vee ha(i',k') - ha(i,k) \geq Se' ,
\end{equation}
where the lower bounds $Se$ and $Se'$ are input data.

Several types of temporal constraints are defined in the original problem definition from ROADEF \cite{roadefWeb}, but they can be converted to the type in \eqref{min_separation}.

For every week $h$ there is a collection of subsets of outages and for each subset $A$ in this collection, an associated natural number $N$. For every $A$ and $N$, at most $N$ of the outages in $A$ are allowed to be on outage in week $h$
\begin{equation} \label{max_offline}
\sum_{(i,k) \in A} \Phi(i,k,h) \leq N ,
\end{equation}
where $\Phi(i,k,h)$ equals $1$ if outage $(i,k)$ is active in week $h$ and $0$ otherwise.

There are limited resources available for carrying out maintenance. Thus, a collection of subsets of outages is given. Each subset $A$ in this collection has an associated resource availability $Q$. For every $A$ and $Q$, at most $Q$ of the outages in $A$ can use resources in any week
\begin{equation} \label{max_resource_usage}
\forall h: \sum_{(i,k) \in A} \phi(i,k,h) \leq Q ,
\end{equation}
where $\phi(i,k,h)$ equals $1$ if outage $(i,k)$ uses resources in week $h$ and $0$ otherwise. Note that the weeks in which an outage uses resources are not necessarily the same as the weeks in which it is active.

Finally, there is a lower bound on the online capacity at a given time. In other words, an upper bound on the capacity that is allowed to be offline at the given time. Thus, a collection of subsets of outages is given. Each subset $C$ in this collection has an associated upper bound $IMAX$ and a subset of weeks $IT$. For every $C$, $IMAX$, and $IT$, during any week in $IT$ the total offline capacity of plants in $C$ cannot exceed $IMAX$
\begin{equation} \label{offline_capacity}
\forall h \in IT: \forall t \in h: \sum_{\substack{i \in C: \exists k : \\ t \in ea(i,k)}} PMAX_i^t \leq IMAX .
\end{equation}
Note that in \eqref{offline_capacity} a week is considered as a set of time steps. The sum is simply over all type 2 plants in $C$ that are offline at time step $t$.

\subsection{Objective function}
The objective function is composed of three terms: the total cost of reloading all type 2 plants, the average cost of type 1 production over all scenarios, and the value of residual fuel at type 2 plants at the end of the planning horizon. Let $C_{i,k}$ denote the cost of fuel for type 2 plant $i$ in cycle $k$, $C_{j,t,s}$ the cost of production for type 1 plant $j$ at time step $t$ in scenario $s$, and $C_i$ the cost of fuel for type 2 plant $i$ at the end of the planning horizon. Then the objective function to be minimized is
\begin{equation} \label{objective_function}
\sum_{i=0}^{I-1} \sum_{k=0}^{K-1} C_{i,k} r(i,k) + \frac{1}{S} \sum_{s=0}^{S-1} \sum_{t=0}^{T-1} \sum_{j=0}^{J-1} C_{j,t,s} p(j,t,s) D - \sum_{s=0}^{S-1} \sum_{i=0} ^{I-1} C_i x(i,T,s) .
\end{equation}

\subsection{NP-hardness} \label{NP-hardness}
To prove the NP-hardness of the problem under consideration, we propose a reduction from 1-in-3-SAT, which is proved to be NP-hard by Schaefer in \cite{schaefer1978}. Reductions directly from a scheduling problem might be possible but is complicated by the exponentially (albeit pseudo polynominal) number of time steps that often will arise.

Given a boolean formula $\beta_1\wedge \dots \wedge \beta_c$ where each clause $\beta_i$, $1 \leq i \leq c$, is the disjunction of three boolean literals (or their negation) from the set $\{x_1 \dots x_n\}$, 1-in-3-SAT asks for an assignment of true or false to $x_1 \dots x_n$ such that exactly one of the literals in each clause $\beta_1\wedge \dots \wedge \beta_c$ evaluates to true.

To solve an instance of 1-in-3-SAT by using the problem under consideration, we construct an instance with a single scenario as follows. A type 2 plant $i$ with a single outage with a duration of one week is created for each clause $\beta_i$. Furthermore, a week is created for each variable $x_h$ and its negation $\neg x_h$, in such a way that the variable and its negation occupy successive weeks. Scheduling an outage in the week that corresponds to $x_h$ (or $\neg x_h$) is interpreted as forcing $x_h$ to be true (or false).

All outages must be scheduled in order for the 1-in-3-SAT instance to be satisfiable. A constraint of type \eqref{outage_bounds} with bounds set to include all $2n$ weeks for every outage ensures this.

To ensure that the single outage for plant $i$ can only be scheduled in one of the three weeks corresponding to literals in $\beta_i$, a constraint of type \eqref{offline_capacity} which restricts the amount of offline capacity is added for the outage. We set $PMAX_i^t = 1$ for all $t$ and $IMAX = 0$. Furthermore, we let $C$ contain plant $i$'s single outage and let $IT$ contain all weeks except the three corresponding to literals in the $\beta_i$. Constraint \eqref{offline_capacity} is enforced on timesteps rather than weeks, so the number of time steps is set to $2n$ such that there is one time step per week.

Constraints of type \eqref{min_separation} are added to ensure that outages are not scheduled in both the week corresponding to $x_h$ and $\neg x_h$. For each pair of clauses that contains the literal $x_h$ and $\neg x_h$ respectively, a constraint of type \eqref{min_separation} is defined on the two weeks corresponding to $x_h$ and $\neg x_h$ (which are consecutive by construction). Forcing a separation of two weeks between the corresponding plants' outages prevents these outages from being scheduled in the weeks corresponding to $x_h$ and $\neg x_h$, respectively.

See Figure \ref{NPexample} for an example of a simple boolean formula encoded as a maintenance scheduling problem.

The construction described above is polynomial in the input size, as we have $c$ clauses, giving rise to $c$ outages, which each has three valid weeks in which it can be scheduled. For each of these weeks, we need less than $c$ constraints to restrict it from conflicting with the other outages. The conversion is thus bounded from above by $3c^2$.


\begin{figure}
\centering
\begin{tabular}{p{0.6cm}|p{0.6cm}|p{0.6cm}|p{0.6cm}|p{0.6cm}|p{0.6cm}|p{0.6cm}|p{0.6cm}}
  $x_1$ & $\neg x_1$ & $x_2$ & $\neg x_2$ &$x_3$ & $\neg x_3$ & $x_4$ & $\neg x_4$  \\
  \hline
  \cellcolor[gray]{0.7} &\cellcolor[gray]{0.3}&&\cellcolor[gray]{0.3}&\cellcolor[gray]{0.3}&&\cellcolor[gray]{0.3}&\cellcolor[gray]{0.3}\\
  \hline
  \cellcolor[gray]{0.3} &\cellcolor[gray]{0.7}&&\cellcolor[gray]{0.3}&\cellcolor[gray]{0.3}&\cellcolor[gray]{0.3}&&\cellcolor[gray]{0.3}\\
\end{tabular}
\caption{A representation of the formula $(x_1\vee x_2 \vee \neg x_3)\wedge(\neg x_1 \vee x_2 \vee x_4)$. There is one row per clause in the formula. Dark gray weeks are disallowed using constraint \eqref{offline_capacity}. Outages can be scheduled in light gray and white weeks, but the light gray weeks for $x_1$ and $\neg x_1$ cannot both be used for outages due to a constraint of type \eqref{min_separation}}
\label{NPexample}
\end{figure}


A single type 1 plant can be used to cover any demand we decide on. We set the demand $DEM^{t,0} = I$ for each time step $t$, and choose $PMAX_i^t = 1$, $PMIN_i^t = 0$ for each type 2 plant $i$ and time step $t$. The initial fuel stock is set large enough to allow every type 2 plant to produce in all $2n$ weeks without outages. Minimum and maximum bounds on refueling of the type 2 plants are set such that they do not constrain the solution, i.e., $RMIN_{i,k}=0$, $RMAX_{i,k}=0$. To ensure that $AMAX_{i,k}$ and $SMAX_{i,k}$ do not become constraining, they are set to the initial fuel stock.

Having scheduled all outages, truth values are assigned to literal in the given 1-in-3-SAT instance as follows. A literal is set to true if some outage is scheduled in the corresponding week and to false otherwise. Thus, any instance of 1-in-3-SAT can be solved by scheduling outages. As menioned above, the size of the reduction's output is polynomial in the size of the 1-in-3-SAT instance and can obviously be constructed in polynomial time, and therefore the optimization problem in this paper is NP-hard in the strong sense, i.e., it is NP-hard even if the numerical parameters are encoded with unary base.

\section{Initial solution construction} \label{initial_solution_construction}
The maintenance scheduling part of the problem is NP-hard, and our experience is that finding non-trivial feasible solutions to this problem is challenging in practice as well. Our initial approach included different directions: an attempted local search, a set of construction heuristics, and a CP approach. The CP approach turned out to be the best approach. Our overall setup thus starts by making a first feasible maintenance schedule using CP.

\subsection{Constraint programming} \label{constraint_programming}
An exact representation of the problem would result in a large number of variables since it requires modelling of every time step. Furthermore, concepts such as modulation, the decreasing power profile, and the cost of type 1 production would have to be included in the model. Instead we focus only on finding a feasible maintenance schedule and introduce a surrogate objective function that approximates the real objective function, thereby leaving the rest of the optimization to the subsequent local search.

A CP model is used to find a feasible maintenance schedule.\footnote{For a general introduction to CP, see the textbook \cite{constraint_programming} by Apt.} For every outage, the CP model has three decision variables in the model: A binary variable $\sigma(i,k)$ deciding if outage $k$ for type 2 plant $i$ is scheduled or not, an integer variable determining the starting week for the outage, and an integer variable determining the refueling level. Refueling levels are continuous in the problem formulation, but are discretized because most CP solvers cannot handle continuous variables. To reduce the domain of the refuel variables in the CP model the discretisization is into segments of 1000 fuel units.

We model the scheduling constraints \eqref{outage_bounds} and \eqref{min_separation}-\eqref{offline_capacity} exactly. But the fuel level constraints \eqref{amax} and \eqref{smax} are approximated because an exact representation requires exact modeling of fuel consumption of type 2 plants, which leads to a too large model.

\subsubsection{Fuel level approximation}
To estimate the fuel level we introduce a set of variables, $FU(i,k)$, which denotes the fuel used during a the production campaign preceeding outage $(i,k)$ assuming maximal production. This is used to calculate $FB(i,k)$ which denotes how much fuel remains before the outage $(i,k)$, and \eqref{fuel_reload} is then used to calculate the amount of fuel that is available in the plant after the outage $(FA(i,k))$. 

To compute these values two additional functions are introduced: Let $\beta(i,h)$ denote the accumulated fuel usage for type 2 plant $i$ in the first $h-1$ weeks of the planning horizon, assuming production at maximum capacity:
\begin{equation} \label{acc_fuel}
\beta(i,h) = D\sum_{h' = 0}^{h-1} \sum_{t \in h'} PMAX_i^t.
\end{equation}
The $\beta(i,h)$ values allows us to easily estimate the fuel usage $FU(i,k)$ during the production period preceeding outage $k$ for type 2 plant $i$:
\begin{equation}
FU(i,k) = \begin{cases} 
             \sigma(i,k) \beta(i, ha(i,k)), & \mbox{if } k=0 \\ 
             \sigma(i,k) (\beta(i, ha(i,k)) - \beta(i, ha(i,k-1)+DA_{i(k-1)})), & \mbox{if } k>0 
          \end{cases}
\end{equation}
An initial estimate $FI(i,k)$ of the fuel level at type 2 plant $i$ the time of outage $k$, which is subsequently adjusted to a better estimate $FB(i,k)$, is obtained as follows
\begin{equation}
FI(i,k) = \begin{cases} 
             XI_i - FU(i,k), & \mbox{if } k=0 \\ 
             FA(i,k-1) - FU(i,k), & \mbox{if } k>0 
          \end{cases}
\end{equation}
where $FA(i,k-1)$ is the estimated fuel level after outage $k-1$ for type 2 plant $i$, computed from $FB(i,k-1)$ and the refuel amount $r(i,k-1)$ using \eqref{fuel_reload}. $XI_i$ is the initial fuel level for plant $i$.

\begin{figure} [t]
	\centering
		\includegraphics[width=\textwidth, trim=0 240 0 70]{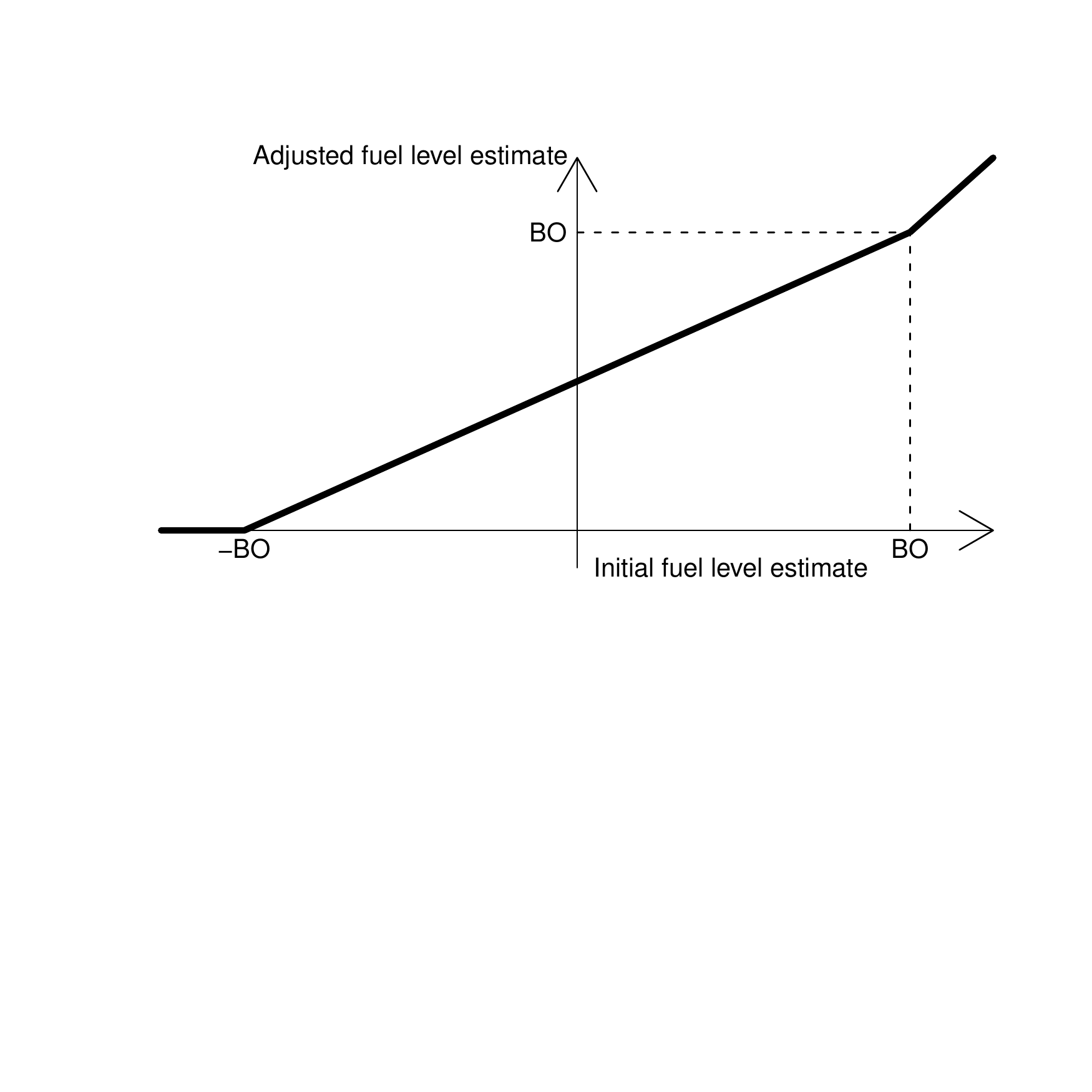}
		\caption{The linear function mapping an initial fuel level estimate $FI(i,k)$ to an adjusted estimate taking the power profile into account. The adjusted estimate is never negative, and for any $FI(i,k) \geq BO$, the function acts like the identity function.}
	\label{fig:fuel_level_estimate}
\end{figure}

The problem with $FI(i,k)$ is that the declining power profile, constraint \eqref{power_profile_enough_fuel}, is ignored and there is no modulation. This will often underestimate the actual fuel level because the plant usually produces at less than $PMAX$ at the end of the production campaign and consequently uses less fuel. Experiments show that it often leads to situations where no feasible solution can be found. Thus, in order to take the declining power profile into account, we adjust $FI(i,k)$ if it is low enough that the power profile is activated in the end of the production campaign. More precisely, if $FI(i,k) < BO_{i,k}$, we assume that $i$ would have run out of fuel when $FI(i,k)=-BO_{i,k}$ and make a linear interpolation, see Figure \ref{fig:fuel_level_estimate}. The decision to choose $-BO_{i,k}$ is heuristic.

The adjusted fuel level estimate $FB(i,k)$ is computed as:
\begin{equation}
FB(i,k) = \sigma(i,k) \max (0, FI(i,k) + \frac{1}{2} \max (0, \min(2BO_{i,k}, BO_{i,k} - FI(i,k) ) ) )
\end{equation}

This relation between $FB(i,k)$ and $FI(i,k)$ is also shown in Figure \ref{fig:fuel_level_estimate}. This approximation implies that a feasible solution might be infeasible in the CP model and vice versa, but in practice the approximated constraints give solutions that are feasible, also in the sense that there exists a feasible production plan for all scenarios.

\subsubsection{Surrogate objective function}
The scheduling problem is primarily concerned with minimizing the use of type 1 plants, which is equivalent to maximizing the amount of available type 2 capacity, so the surrogate objective function is to minimize the average offline type 2 capacity. A formal specification of the objective function requires some notation. Let $\alpha_i$ denote the average maximal production per week for type 2 plant $i$:
\begin{equation}
\alpha_i = \frac{\sum_{t=0}^{T-1} PMAX_i^t}{H} D.
\end{equation}
Furthermore, the auxiliary decision variable $k'_i$ denotes the index of the last scheduled outage for type 2 plant $i$:
\begin{equation}
k'_i = \sum_{k=0}^{K-1} \sigma(i,k) - 1.
\end{equation}

Then, the surrogate objective function to be minimized is:
\begin{align}
\displaystyle \nonumber
   \sum_{i=0}^{I-1} \alpha_i \bigg( & \max \big(0, ha(i,0) - \frac{XI_i}{\alpha_i} \big) + \\ 
\displaystyle \nonumber
&  \sum_{k=1}^{K-1} \sigma(i,k) \max \big(0, ha(i,k)-(ha(i,k-1)+DA_{i(k-1)}) - \frac{FA(i,k-1)}{\alpha_i} \big) \\
\displaystyle \nonumber
&  + \max \big(0, H - (ha(i,k'_i) + DA_{ik'_i}) - \frac{FA(i,k'_i)}{\alpha_i} \big) \bigg)
\end{align}
The first term in the surrogate objective function is the estimated offline capacity before the plant's first outage, the second term is estimated offline capacity before each of the plant's subsequent outages, and the third term is the estimated offline capacity after the plant's last outage.

\subsubsection{Search strategy}
A CP solver finds a solution to an optimization problem by searching a tree which is pruned by applying \emph{constraint propagation} and \emph{branch and bound} strategies. Two decisions are crucial for making this pruning effective, namely \emph{variable selection} (choosing the next variable to branch on) and \emph{value selection} (choosing a value for the chosen variable). Variable and value selection strategies are generally chosen according to the \emph{first-fail} principle, which says that if no feasible solution exists, then the search should determine this as early as possible. Furthermore, finding a good solution early in the search is desirable because it improves the efficiency of branch and bound pruning.

Our variable selection strategy is to make decisions that concern outages which are scheduled closely as close together in the search tree as possible. This is achieved by branching on variables grouped by outage in the following way. The plants are randomly permutated, and index $i$ then corresponds to index $\rho(i)$ after the permutation. We go through all cycles $k \in \{0, \dots , K-1\}$, and for each $k$ through all type 2 plants $i \in \{0, \dots , I-1\}$. For each outage $\rho(i),k$ the variables are fixed in the following order and with the specified value selection strategy:
\begin{description}
	\item Determine whether $(\rho(i),k)$ is scheduled, i.e., whether $\sigma(\rho(i),k) = 1$ or $0$. The $\sigma(\rho(i),k) = 1$ ie. $ha(\rho(i),k) > -1$ branch is considered first, since this leads to more scheduled outages.
	\item Determine the starting week $ha(\rho(i),k)$. The earliest possible week is considered first, since this leaves more room for subsequent outages for plant $\rho(i)$.
	\item Determine the refuel amount $r(\rho(i),k)$. The maximal amount is considered first, since this leads to more type 2 capacity.
\end{description}
Preliminary experiments showed that this branching strategy works well. Details about the CP solver softwre are given in Section \ref{implementation_details}.

\subsection{Aggregating scenarios} \label{sec:aggregate}
The solution found by the CP solver specifies starting time and refueling amount of each outage but no production levels since the latter are scenario specific. The large number of scenarios with different demands and type 1 costs complicates computations. In all feasibility computations that come after the CP solver we therefore use a minimum demand scenario where $DEM^{t}_{min} = \min_s (DEM^{t,s})$ for all $t$. By using the minimum demand scenario we ensure feasibility of all scenarios while only checking a single one.

The only way the demand influences feasibility is by making modulation necessary to get the type 2 production low enough to match demand in constraint \eqref{demand_constraint}. As type 2 powerplants according to \cite{roadefWeb} delivers an average of 87\% of the combined power, such situations arise sparingly, and the minimum demand scenario can be modulated to feasibility in all instances used in the competition.

When evaluating the objective function, an average type 1 cost over all scenarios is used. If total type 2 production does not cover the demand in some time step, the uncovered demand is met using type 1 plants that have a fixed cost per unit of power produced. This gives rise to a piecewise linear function for the cost in each scenario, mapping type 2 production to the cost of covering the slack using type 1 plants. These piecewise linear functions are then aggregated into a new piecewise linear function that maps type 2 production to type 1 cost of covering the slack averaged over all scenarios. The new function will have up to $J\cdot S$ breakpoints, which may be a large number. We explain how we cope with this issue in in Section \ref{type1cost}.

\subsection{Greedy production level planning}
\label{sec:greedy}
We set the production levels $p(i,t,s)$ and refuel amount $r(i,k)$ of a feasible solution returned by the CP solver by means of a greedy algorithm which we call production planner. Only constraint \eqref{demand_constraint} concerning demand binds production across different type 2 plants. The production planner ignores the demand and therefore all computations can be done for each type 2 plant independently. Ignoring the demand may lead to overproduction which is fixed by modulation in a final phase which is described in Section \ref{modulation}.

The algorithm starts with the first time step and goes through all time steps in the schedule. It uses the initial fuel level to produce at maximum capacity until no more fuel remains or the next outage is encountered. If a plant runs out of fuel in some production campaign, it cannot produce in the rest of the production campaign.

We use the production planner in two settings: it is used with an initial maintenance schedule from the CP solver, and later it is used repeatedly in the local search. When applied to a solution from the CP solver the production planner initially sets refuel amounts to the minimum allowed amount $RMIN_{i,k}$ for every outage. When called from local search the current refuel amounts are reused. Subsequently, the production planner first tries to achieve feasibility, see the next section, and then to increase the refueling amount as described in Section \ref{sec:increase_refuel}.


\subsubsection{Reducing refuel amounts}
Infeasibility with respect to fuel levels can occur if constraint \eqref{amax} is violated because of a too high fuel level before an outage, or constraint \eqref{smax} is violated because of a too high fuel level after an outage. If either of these situations is encountered the production planner backtracks to the previous outage and reduces the amount of refueling done there - this change is subject to constraint \eqref{reload_bounds}. This is done recursively, and if the backtracking reaches the start of the planning horizon without resolving the problem, the planner declares the maintenance schedule infeasible. If this happens and the production planner is called from the local search, the infeasible neighbor is simply skipped. It has never happened to the initial solution from the CP solver, but in this case our algorithm is unable to solve the given instance.


\subsubsection{Increasing refuel amounts}
\label{sec:increase_refuel}
After having decreased refueling amounts wherever necessary to the point where constraints \eqref{amax} and \eqref{smax} are satisfied, we try to increase the refuel amounts as much as possible in order to maximize type 2 production. This is done for each production campaign $k$ in turn, starting with the last. If plant $i$ is able to produce at $PMAX_i^t$ for all $t \in ec(i,k')$ where $k'\geq k$, then there is no need to increase refuel amounts. Otherwise, we try to increase the refueling amount in 2\% increments of the difference between the maximum allowed amount $RMAX_{i,k}$ and the current refuel amount $r(i,k)$.

The production planner is used to evaluate feasibility, and if infeasibility is detected, the increment of refuel amount is undone and the next production campaign is considered.

\section{Improvement by local search} \label{improvement_by_local_search}
The first complete solution obtained by applying greedy production planning to the initial CP scheduling solution is of relatively low quality, since the strategy used in the CP model is to place outages mainly to ensure a feasible solution and to a lesser extent to minimize production costs. This leaves room for improving the temporal placement of outages.

Moving outages can reduce the total cost of production in two ways: first, it can increase the amount of electricity produced by type 2 plants and thereby decrease the production of the more expensive type 1 plants. Second, it can move outages to a time period where the alternative type 1 production costs are low.

A limitation of the following local search procedure is that the number of outages is not changed. The local search only tries to reoptimize the placement of the already scheduled outages, not remove outages or introduce new ones. The effect of this is that the CP model is trusted with deciding the number of outages for each power plant. 

The basic idea in the local search is to choose a random outage and move it a few weeks forward or backward. After each move the scheduling constraints \eqref{outage_bounds} and \eqref{min_separation} to \eqref{offline_capacity} are checked for feasibility. If the constraints are satisfied the production planner calculates updated refueling amounts and production levels in order to check feasibility with respect to fuel levels. If a move is feasible, the change in cost must be evaluated in an efficient way in order for the search to visit a large number of solutions.

\subsection{Neighborhood}
Formally, given a schedule for outages, $ha(i,k)$, $i \in I, k \in K$, a neighboring scheduling solution $ha'$ obtained by applying the move $(i',k',m)$ is:
\begin{equation}
ha'(i,k) = \left\{
                  \begin{array}{ll}
                     m,       &  \mbox{ if } i=i' \mbox{ and } k=k' \\
                     ha(i,k), &  \mbox{ otherwise} 
                  \end{array}
           \right.
\label{eq:ls_neighbor_sol}
\end{equation}
The value $m$ is chosen in the interval $\left[  TO_{i',k'}, TA_{i',k'}\right]$, so only neighboring schedules that satisfy constraint \eqref{outage_bounds} are considered. A move $(i',k',m)$ corresponds to selecting outage $k'$ of plant $i'$ and moving it to start in week $m$. 

The size of the neighborhood is bounded from above by $I \cdot K \cdot H$, but the bounds in \eqref{outage_bounds} reduce the number of neighbors significantly. The length of the interval $[TO_{i,k}, TA_{i,k}]$ is usually between 20 and 30 weeks in average (including outages where $TO_{i,k}$ or $TA_{i,k}$ are undefined, in which case the interval is everything to the left of $TA_{i,k}$ or everything to the right of $TO_{i,k}$). In two instances where very few outages have constraints of this type, the average is around 150 weeks. The size of the neighborhood is further reduced by only considering moves $(i,k,m)$ where the outage is moved less than $n$ weeks forward or backward, i.e.,
\begin{equation}
  \left| ha_{i,k} - m \right| < n.
\label{eq:ls_neighborhood_decrease}
\end{equation}
Experiments have shown that a good value for $n$ is 20. The feasibility of a neighbor can be checked effectively because each outage is involved in a relatively low number of constraints. It is straightforward to precompute a matrix that maps an outage to its corresponding set of scheduling constraints, and this matrix can be used to check if some constraint in this set is violated after the starting time of the outage has been changed. If the feasibility check detects a violated constraint, the evaluation is terminated immediately and the move is discarded. This means that the local search never moves to an infeasible maintenance schedule.

\subsection{Delta evaluation} \label{delta_evaluation}
Calculating the change in solution cost is more complicated. First it is necessary to replan the production levels for the type 2 power plant that had an outage moved. As described in Section \ref{sec:aggregate} only a single scenario of production levels for each type 2 plant is maintained. The production levels and implied fuel levels must be recalculated after an outage has been moved to ensure that we still have a feasible production plan. The new production plan is calculated by the production planner described in Section \ref{sec:greedy}. If the production planning fails, the move is discarded.




After the greedy production planner has calculated new production levels, the delta value $\Delta = \Delta_{refuel} + \Delta_{type1} - \Delta_{remainder}$ is calculated, where $\Delta_{refuel}$ is the change in cost of type 2 refueling, $\Delta_{type1}$ is the change in cost of type 1 production, and $\Delta_{remainder}$ is the change in value of remaining fuel at the end of the planning horizon. Computation of these three numbers is described below.

The change in refuel cost is 
\begin{equation}
\Delta_{refuel} = \sum_{k=0}^{K-1}{C_{i,k} (r'(i',k) - r(i',k))} ,
\end{equation}
where $r'(i',k)$ is the refuel amount in the neighboring solution.

The change in remaining fuel at the end of the planning horizon can be estimated. This can vary from scenario to scenario, but here we estimate the cost using a special scenario $s^*$ in which every type 2 plant has full production
\begin{equation}
\Delta_{remainder} = S \cdot C_i( x'(i,T,s^*) - x(i,T,s^*)) ,
\end{equation}
where $x'(i,T,s)$ is the fuel level in the neighboring solution. This estimate is therefore a lower bound on the actual value. 

The change in total cost of type 1 production $\Delta_{type1}$ is more complicated and described in Section \ref{type1cost}.

\subsection{Estimating cost of type 1 production} \label{type1cost}
To calculate $\Delta_{type1}$, we consider the change in production by the moved type 2 plant in each time step, and from this update a list of total type 2 production for each time step. In scenarios where the demand is higher than the total type 2 production, the difference must be covered by type 1 plants.

The type 1 plants can be preordered by increasing production costs, and thus the exact change in total type 1 cost given a change of total type 2 production can be computed in $O(J)$ time for a single time step and scenario. Summing this exact change over all time steps and scenarios is of complexity $O(T \cdot S \cdot J)$ which is too high for local search. Thus, we approximate the change in cost. The approximation removes the need to consider all scenarios and the list of type 1 plants but not the need to consider all time steps, and thus the complexity of the approximation is $O(T)$. 

\begin{figure} [tb]
  \begin{center}
    \includegraphics[scale=0.8, trim=0 30 0 0]{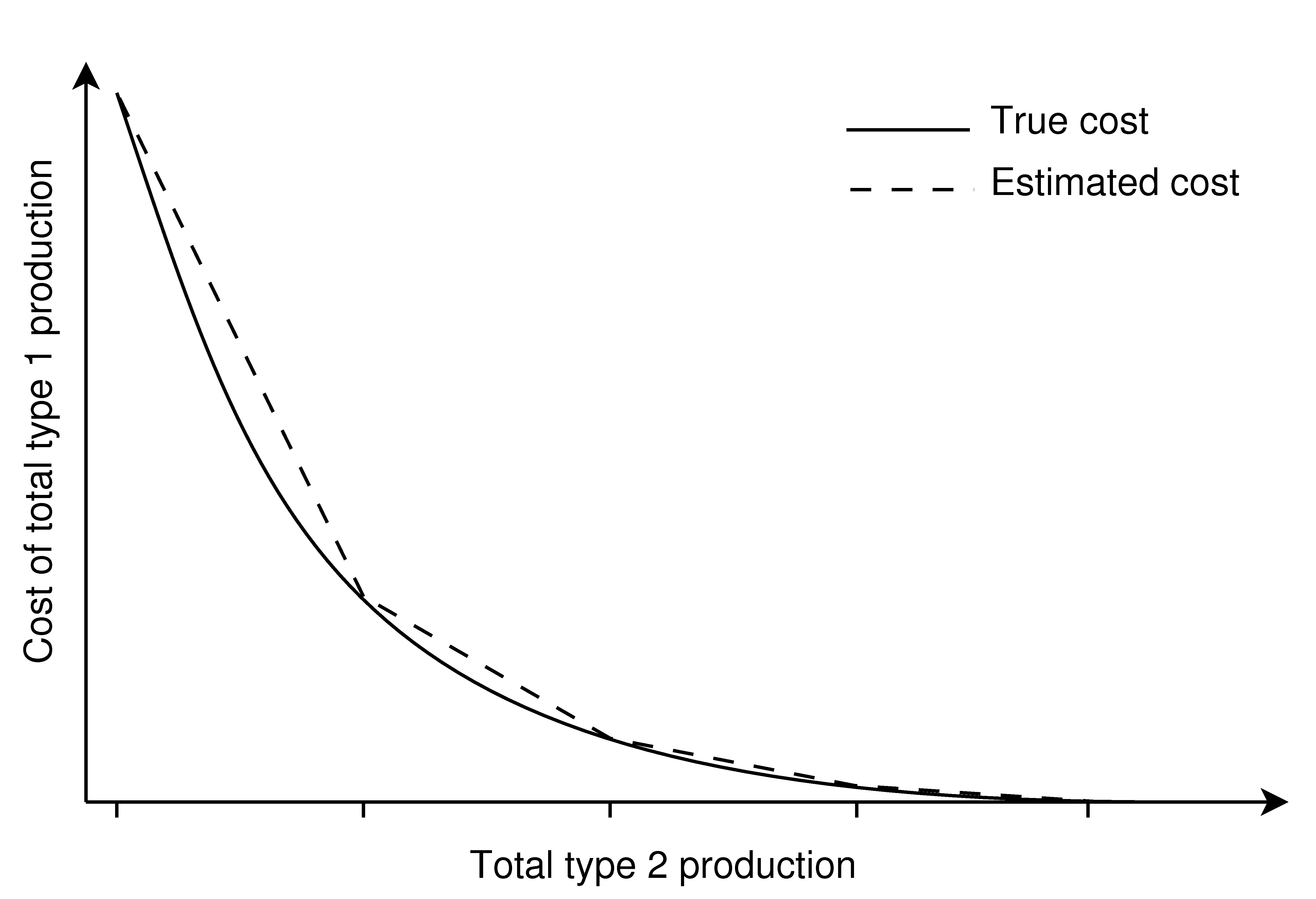}%
  \end{center}
\caption{Linear approximation of total type 1 costs for a single time step $t$}%
\label{fig:ls_exact}%
\end{figure}

To estimate the change in type 1 cost for a single time step, we precompute a piecewise linear function which maps total type 2 production to total type 1 cost for each time step. The solid line in Figure \ref{fig:ls_exact} shows an example of such a function. The function is shown smooth but is actually a piecewise linear function with many breakpoints. As the total type 2 production increases the need for type 1 production diminishes, and when total type 2 production reaches the maximum demand over all scenarios, the total type 1 cost is zero.

The many breakpoints make evaluation of the function computationally expensive because one must go through the breakpoints in order to find the interval containing the current total type 2 production (a binary search speeds up this proces but is still too slow). Therefore we approximate it with another precomputed piecewise linear function with fewer breakpoints which are chosen on the x-axis in an equidistant way. The dashed line in Figure \ref{fig:ls_exact} shows an example of this approximation. Since the actual type 1 cost is convex, the approximation is an upper bound on the actual cost. Experiments show that this approximation is relatively good: For $3 \cdot I$ breakpoints the approximation error has been sampled on all instances, and if absolute type 1 production demands under 100 units are ignored, the worst average deviation over an instance was $\approx 3.36\cdot10^{-4}$ percent per measurement. This is significantly less than the differences in cost encounted during local search.

Using the precomputed approximation we can perform a constant time evaluation of the total type 1 cost for a certain total type 2 production in a single time step. No search for the stored breakpoint closest to a given total type 2 production is required, because the breakpoints are equidistant. If the total type 2 production falls between two stored breakpoints, we use linear interpolation to get the total type 1 cost. Formally, if $F(t,i)$ is the cost of type 1 production in time step $t$ at breakpoint $i$, the interpolation is done between the indices:
\begin{align}
	i_{low} = \left\lfloor \frac{P_2^{total}}{Int} \right\rfloor, \qquad i_{high} = \left\lceil \frac{P_2^{total}}{Int} \right\rceil
\label{eq:constant_time_eval}
\end{align}
where $P^{total}_2$ is the sum of type 2 production in time step $t$, and $Int$ is the size of the interval between two breakpoints.

Despite this approximation of the change in total type 1 cost, the evaluation is slow compared to the scheduling constraint violation check because it must be computed for every time step. However, the evaluation of changes in cost is only performed if all the scheduling constraints are satisfied.

\subsection{Simulated annealing}
A typical simulated annealing metaheuristic is used to guide the local search. For a general introduction to simulated annealing in optimization, see e.g. Kirkpatrick ét al  \cite{kirkpatrick1983optimization}. The simulated annealing algorithm accepts a move with change in cost $\Delta_{c}$ with probability
\begin{equation}
  p = \min \left(1, \exp\left(-\frac{\Delta_{c}}{\tau}\right)\right),
\label{eq:sa_prob}
\end{equation}
where $\tau$ is the current temperature. This implies that an improving move is always accepted since $\exp(x) > 1$ for any positive $x$.

The initial temperature is dynamically set such that about half of the considered moves are accepted at the beginning of the search. This is done by evaluating neighbors of the initial solution and adjusting the temperature until half of these neighbors are accepted. The cooling scheme is geometric with plateaus (sometimes called piecewise constant), which means that the temperature is not lowered in every iteration. Johnson et al.\ (see \cite{Johnson89} and \cite{Johnson91}) show experimentally that plateaus in the cooling scheme lead to better solutions when using simulated annealing to solve graph partitioning and graph coloring problems. The cooling ratio $c$ and the number of moves $n_{plateau}$ at each temperature are given as parameters. When $m_{idle}$ moves in a row have been considered without any move being accepted, a restart is performed. When the search is restarted, the current temperature is set to $k_{restart}$ times the starting temperature of the previous annealing run. This is done to help the local search move away from previous local optima.

\section{Modulation} \label{modulation}
The local search creates a solution where the cost of covering demand by type~1 power plants is minimized. This may lead to solutions with overproduction at the type 2 plants in some time step $t$ in some scenario $s$:
\begin{equation*}
\sum_{i=0}^{I-1} p(i,t,s) > DEM^{t,s} .
\end{equation*}
Such overproduction can be eliminated in two ways: modulation can be used to decrease the power output of a type 2 plant in the affected time steps, or refueling can be decreased such that a type 2 plant runs out of fuel before that time step. Making a power plant run out of fuel will limit the plant's fuel level in the next campaign. Furthermore, it eliminates the rest of its production for the rest of the campaign in all scenarios, since refuel amounts are shared among all scenarios. This method thus becomes a last resort, i.e., modulation is used whenever possible and is capable of eliminating overproduction on all the examined instances.

There are two constraints on the amount of modulation that can be performed on a single power plant: constraint \eqref{eq:maxmod} restricts the amount of modulation that can be performed in the current campaign, and constraints \eqref{amax} and \eqref{smax} enforce an upper limit on the fuel level before and after refueling. The latter can be handled by adjusting the refueling of the power plant, but this will affect all scenarios.

Since refuel amounts for a plant are shared among all scenarios, we use a two-step procedure to eliminate overproduction. First, we ensure that there exists a refueling scheme that is feasible for all scenarios using the minimum demand scenario defined in Section \ref{sec:aggregate}. Second, we determine modulation for each individual scenario.

\subsection{Modulation for the minimum demand scenario}
A modulation and refueling scheme is created for the minimum demand scenario, thus ensuring that the refueling is feasible for all scenarios. This is done by running through the time steps in increasing order. When a time step with overproduction is detected, a target plant $i$ is selected among the type 2 plants. Plant $i$ has its production lowered as much as possible with respect to the amount of remaining modulation capacity in that campaign as given in equation \eqref{eq:maxmod}. Subsequently we repair refuel values on plant $i$ using the greedy production level planner on plant $i$. If this fails, the modulation on plant $i$ is undone, and another plant is selected.

As the cost of modulating each type 2 plant is the same, modulation can be seen as an available resource, which expires when a production campaign ends. Thus, the target plant selection strategy iterates through plants in non-descending order of their current campaign's end date.

A refuel plan may be infeasible for the minimum demand scenario but still be feasible for all scenarios. Our method is unable to cope with this situation and will therefore declare a schedule infeasible, but it never happened in the instances of the competition.


\subsection{Modulation per scenario}\label{sec:modulationPerScenario}
After completing modulation for the minimum demand scenario, we fix refuel amounts and apply the same modulation algorithm on each scenario. In some cases this step will decrease the objective value by more than 1\%. 

\section{Computational analysis and results} \label{computational_analysis}
In this Section we describe the problem instances used for computational tests, tuning of the parameters in the simulated annealing algorithm, how much time is spent in different components of the heuristic, and finally the results we obtained.

\subsection{Problem instances}
We have tested our algorithm on ten real-life instances supplied by \'{E}lectricit\'{e} de France. Table \ref{statistics} shows various figures as well as the best known objective value for each of the instances, which are taken from the ROADEF website \cite{roadefWeb}. Note that the $B$ and $X$ instances are pairwise very similar, e.g. $B6$ and $X11$, $B7$ and $X12$ and so on. This similarity is due to the fact that they are based on the same data, but they differ in that different filters have been applied to the demand.
\begin{table} [htb] 
\center
\begin{tabular} {c c c c c c c c} \hline
Instance  &  File size  &  T                &  Weeks  &  S    &  J   &  I   &  Best solution          \\ \hline
B6        &  140        &  \numprint{5817}  &  277    &  50   &  25  &  50  & \numprint{83424716217}  \\
B7        &  144        &  \numprint{5565}  &  265    &  50   &  27  &  48  & \numprint{81174243138}  \\
B8        &  262        &  \numprint{5817}  &  277    &  121  &  19  &  56  & \numprint{81926206073}  \\
B9        &  262        &  \numprint{5817}  &  277    &  121  &  19  &  56  & \numprint{81750858197}  \\
B10       &  252        &  \numprint{5565}  &  265    &  121  &  19  &  56  & \numprint{77767024999}  \\
X11       &  140        &  \numprint{5817}  &  277    &  50   &  25  &  50  & \numprint{79116772289}  \\
X12       &  143        &  \numprint{5523}  &  263    &  50   &  27  &  48  & \numprint{77589910940}  \\
X13       &  262        &  \numprint{5817}  &  277    &  121  &  19  &  56  & \numprint{76449207715}  \\
X14       &  262        &  \numprint{5817}  &  277    &  121  &  19  &  56  & \numprint{76172998633}  \\
X15       &  250        &  \numprint{5523}  &  263    &  121  &  19  &  56  & \numprint{75101398439}  \\ \hline
\end{tabular}
\caption{Overview of the ten instances showing for each instance file size in megabytes, number of time steps (T), number of weeks, number of scenarios (S), number of type 1 plants (J), number of type 2 plants (I), and the objective value of the best known solution.}
\label{statistics}
\end{table}

\subsection{Tuning the simulated annealing}
To achieve a set of parameters that perform well for all problem instances we performed the following comparison of different parameter settings. All the following combinations of parameters were run for all ten instances in sets B and X and for ten different random seeds. In order to reduce the number of configurations, we decided after preliminary testing to fix the number of moves at each tempature to 100 and that the initial temperature after a restart should be twice the initial temperature of the previous run.
\begin{description}
\item Cooling ratio $c$ = 0.95, 0.96, 0.97, 0.975, 0.98, 0.985, 0.99, 0.995
\item Start acceptance ratio = 0.25, 0.5, 0.75
\item Stop criterium = 25, 50, 75, 100, 125, 150, 175, 200, 300, 500, 800
\item Number of moves per temperature plateau $n_{plateau}$ =  100
\item Reheat constant $k_{restart}$ = 2
\item Number of moves without acceptance $m_{idle}$ = 50
\end{description}

The highest average solution quality comes from setting the cooling ratio to 0.995, the start acceptance ratio to 0.5 and the stop criterium to 125 non-accepted neighbors in a row. This setting also have a reasonably low variance compared to other settings.

\subsection{Implementation details} \label{implementation_details}
The algorithms are implemented in Java, and the scheduling problem is solved using the Gecode CP solver (see \cite{gecode}). The version of our program that was submitted for the qualifying phase used ILOG's CP solver instead of Gecode, but the former was unable to solve the large problem instances used in the final round. The Gecode solver allows the user large control over the applied branching strategy, and the strategy described in Section \ref{constraint_programming} was used to find a feasible scheduling solution after less than two minutes in all instances, as seen in Table \ref{cpsolver}.

\begin{small}
\begin{table} [htb]\footnotesize 
\center
\begin{tabular} {c c c c c c c c c c c} \hline
	&B6	&B7	&B8	&B9	&B10	&X11	&X12	&X13	&X14	&X15\\ \hline
CP (s)	&84	&67	&11	&22	&28	&13	&8	&13	&9	&15\\
Mod (s)	&7	&7	&48	&43	&43	&8	&10	&70	&71	&49\\
Total (s)&167	&153	&256	&237	&263	&108	&94	&315	&366	&242\\
\% gap	&12.17	&6.73	&16.26	&19.79	&9.80	&9.09	&5.80	&14.97	&15.52	&7.37\\ \hline
\end{tabular}
\caption{The time in seconds needed to produce the first feasible - but not returned - solution (CP), modulating it (Mod) and total time including reading from and writing to disk (Total). The last row is the percentwise gap from best known objective value.}
\label{cpsolver}
\end{table}
\end{small}

On the basis of preliminary observations, we decided to stop the CP solver after ten minutes of the total time available and then return the best solution found. If no feasible solution has been found after ten minutes, we let the solver run until the first feasible solution is found. Letting the solver continue after having found a feasible solution results in more scheduled outages for each type 2 plant, which is important since the subsequent local search does not change the number of scheduled outages. Preliminary tests indicated that ten minutes for the CP solver and the remaining 50 minutes for local search and other tasks is a reasonable distribution of the one hour available. Letting the CP solver run longer in order to further improve the surrogate objective function is not the best time utilization because the correlation between the surrogate and real objective function is not that strong.

\begin{figure} [tb]
  \begin{center}
    \includegraphics[scale=0.4, trim=0 160 0 90]{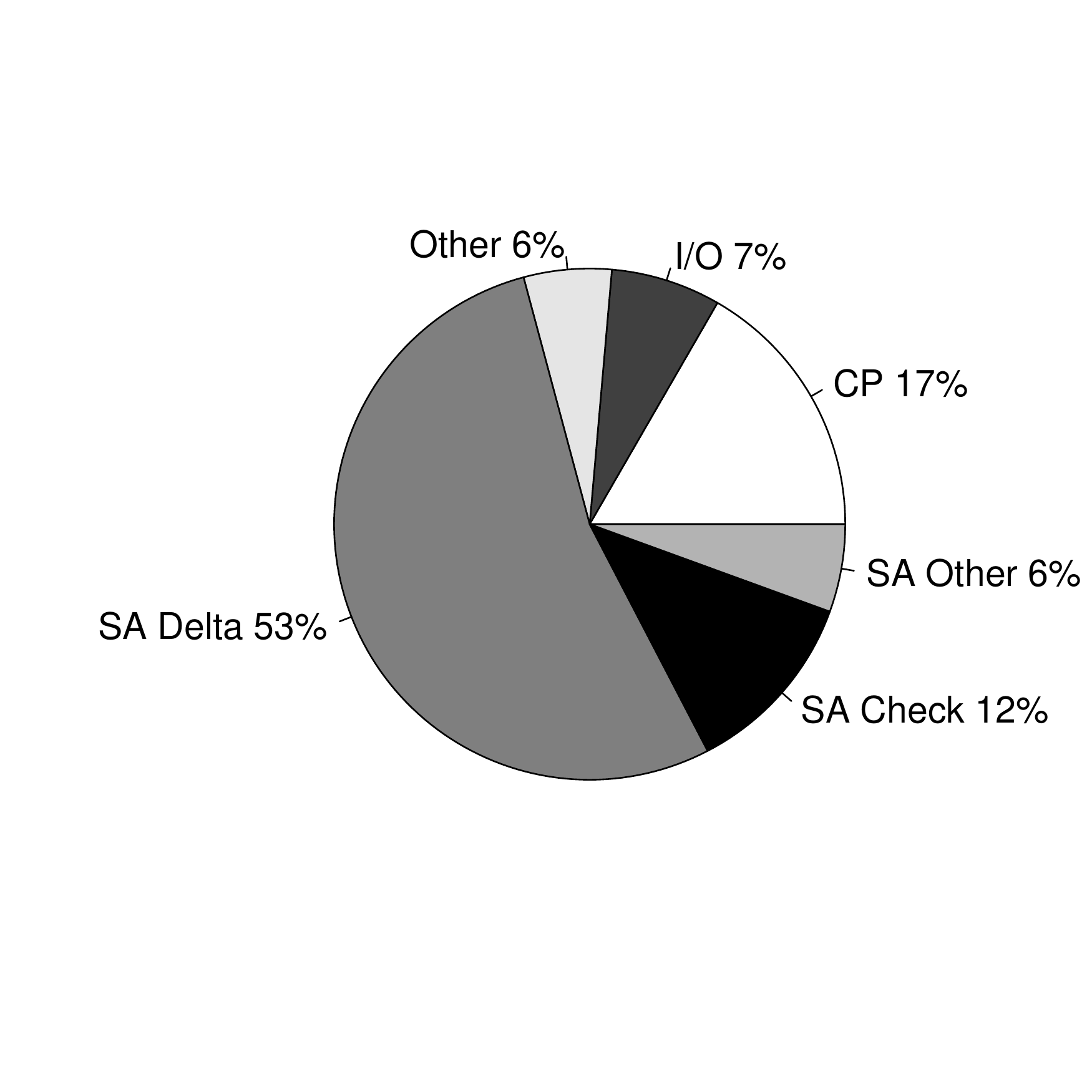}
  \end{center}
\caption{How one hour of wall clock time is spend when solving instance B10.}
\label{fig:pie_chart}
\end{figure}

The pie chart in Figure \ref{fig:pie_chart} shows how much time is usually spent in different parts of the program. Not surprisingly, most time is spent on the simulated annealing algorithm, whose delta evaluation alone accounts for more than half of the total running time. This is due to the fact that evaluation of a neighbor requires replanning of production levels which is very time consuming, even when applying the approximation described in Section \ref{delta_evaluation}. Somewhat unusual is the 7\% of the total running time spent on reading an instance from and writing a solution to the harddisk, which is caused by the very large instance and solution files. The latter takes up to 950 megabytes of harddisk space.

\subsection{Results}
Table \ref{competition_results} shows, for all teams participating in the competition, percentage wise deviation from best known solution for each instance. The last column shows the teams' final score which determined the outcome of the competition. This score is the sum of all ten percentages. If a team was unable to find a feasible solution for an instance, their score for this instance was set to twice the objective value of the worst found solution.

Our team id is J06 which is ranked seven in the table. The first row in the table shows the objective values obtained by our program after fixing the mentioned bug. The results from the table indicate that our new corrected program would win the competition, however note that team S24 would win the competition if they are able to correct their program so it also solves instance X15 to similar quality.

The difference in solution quality between our corrected and uncorrected version of the algorithm, is mainly due to the use of modulation per scenario described in Section \ref{sec:modulationPerScenario}.
\begin{sidewaystable} [ht!] 
\center
\begin{tabular}{ c c c c c c c c c c c c } \hline 
Team  &  B6     &  B7     &  B8       &  B9       &  B10    &  X11    &  X12     &  X13     &  X14     &  X15     &  Score    \\ \hline
J06 (corrected)  & 2.45 	& 0.81 	  & 1.04 	  & 1.23 	  & 1.66 	& 1.30 	  & 1.00 	 & 1.44 	& 1.70 	   &	0.43    & 13.04  \\ \\
S21   &  4.14   &  3.09   &  6.30     &  6.29     &  3.70   &  3.13   &  1.58    &  5.41    &  4.29    &  1.28    &  39.21    \\
J08   &  3.85   &  6.14   &  16.94    &  23.83    &  9.19   &  2.85   &  2.66    &  2.48    &  5.15    &  4.03    &  77.12    \\
S14   &  11.55  &  9.48   &  13.13    &  25.82    &  12.66  &  10.37  &  11.13   &  12.85   &  18.03   &  14.88   &  139.89   \\
S23MT &  1.54   &  0.82   &  1.28     &  2.29     &  1.24   &  TO     &  0.78    &  INF     &  TO      &  \bf{0}  &  202.36   \\
S17   &  19.44  &  25.17  &  39.71    &  56.68    &  34.47  &  16.82  &  23.14   &  TO      &  20.87   &  20.53   &  297.19   \\
S24   &  \bf{0} &  0.14   &  \bf{0}   &  1.06     &  \bf{0} &  \bf{0} &  0.06    &  \bf{0}  &  \bf{0}  &  INF     &  299.96   \\
\emph{J06}   &  \emph{2.50}   &  \emph{0.96}   &  \emph{1.94}     &  \emph{2.12}     &  \emph{3.11}   &  \emph{1.46}   &  \emph{1.11}    &  \emph{2.91}    &  \emph{INF}     &  \emph{INF}     &  \emph{407.5}    \\
S04   &  7.68   &  6.41   &  15.64    &  26.80    &  8.58   &  TO     &  6.50    &  INF     &  14.01   &  TO      &  486.04   \\
S22   &  0.59   &  0.06   &  0.18     &  \bf{0}   &  0.33   &  INF    &  0.01    &  TO      &  ERR     &  ERR     &  494.29   \\
S08   &  INF    &  17.13  &  24.90    &  24.68    &  40.22  &  10.20  &  6.50    &  20.18   &  13.85   &  INF     &  495.24   \\
J05   &  3.34   &  3.43   &  339.67   &  162.91   &  22.43  &  3.79   &  1.52    &  3.35    &  14.35   &  19.41   &  574.20   \\
S23   &  1.54   &  0.82   &  1.33     &  2.28     &  1.30   &  TO     &  TO      &  INF     &  TO      &  TO      &  624.03   \\
S16   &  3.44   &  INF    &  87.05    &  47.31    &  7.15   &  INF    &  2.59    &  TO      &  TO      &  INF     &  773.51   \\
S10   &  7.39   &  66.42  &  3685.91  &  4779.61  &  93.76  &  17.94  &  15.98   &  TO      &  46.34   &  48.21   &  8801.92  \\
S10MT &  7.49   &  66.42  &  3685.91  &  4779.61  &  93.76  &  30.69  &  61.82   &  TO      &  46.34   &  149.35  &  8961.75  \\
J16   &  11.10  &  12.66  &  TO       &  1845.78  &  12.20  &  7.70   &  8.09    &  TO      &  12.81   &  8.11    &  9330.63  \\
S11   &  9.47   &  6.26   &  1902.45  &  MEM      &  INF    &  6.52   &  7.41    &  MEM     &  11.42   &  INF     &  12029.32 \\
S22MT &  0.60   &  \bf{0} &  INF      &  INF      &  INF    &  INF    &  \bf{0}  &  TO      &  ERR     &  ERR     &  17612.27 \\
S25   &  CRA    &  CRA    &  CRA      &  CRA      &  CRA    &  CRA    &  CRA     &  CRA     &  CRA     &  CRA     &  19907.04 \\ \hline
\end{tabular}
\caption{Percentage wise deviation from best known solution for all teams participating in the competition as well as our improved program. Our team is J06. \emph{INF} means that the found solution is infeasible. \emph{CRA} means that the program crashed. \emph{MEM} means that the program ran out of memory. \emph{ERR} means that the format of the solution is invalid. \emph{TO} means that the program did not finish in time.}
\label{competition_results}
\end{sidewaystable}

\section{Conclusion and future work} \label{conclusion}
We have described our approach for solving a large-scale real-life optimization problem using a combination of CP and greedy and local search heuristics. A proof of the NP-hardness of the problem is also given. The solutions obtained are competitive when compared to those found by other teams participating in the ROADEF/EURO Challenge 2010. Other teams achieve better solutions, but the difference between the best known solution and our solution is in the worst case 2.45\% on all instances used in the competition. After fixing the implementation bug in the modulation procedure, our approach is robust in the sense that it is always able to find a feasible solution.

Our approach can be extended in different ways. One possibility is a new local search neighborhood relation which is allowed to change the number of scheduled outages for a type 2 plant, as opposed to the currently used neighborhood which only moves outages. However, it is not easy to fit in additional outages without violating the temporal constraints (constraint \eqref{min_separation}), so the local search would probably need to go through a number of infeasible solutions after adding an outage. Another possibility is to modulate when type 1 production is cheap in order to save fuel which can then be used in subsequent time steps where type 1 production is more expensive.

\paragraph{Acknowledgments}
We would like to thank Marco Chiarandini and Mette Gamst for reading early versions of this paper and providing numerous suggestions for improvement and clarification.
\clearpage

\bibliographystyle{plain}
\bibliography{roadef}

\end{document}